\documentclass[prl,twocolumn,showpacs,preprintnumbers,amsmath,amssymb]{revtex4}

\usepackage{graphicx}
\usepackage{dcolumn}
\usepackage{bm}

\begin{document}

\newcommand{\ord}{LaBaMn$_2$O$_6$}
\newcommand{\disord}{La$_{0.5}$Ba$_{0.5}$MnO$_3$}
\preprint{APS/123-QED}

\title{Phase Separation in A-site Ordered Perovskite Manganite LaBaMn$_2$O$_6$ Probed by $^{139}$La and $^{55}$Mn NMR}

\author{Y.~Kawasaki, T.~Minami, Y.~Kishimoto, T.~Ohno}
\address{Department of Physics, Faculty of Engineering, Tokushima University, Tokushima 770-8506, Japan}
\author{K.~Zenmyo, H.~Kubo}
\address{Faculty of Engineering, Fukuoka Institute of Technology, Fukuoka 811-0295, Japan}
\author{T.~Nakajima, and Y.~Ueda}
\address{Institute for Solid State Physics, University of Tokyo, Kashiwa 277-8581, Japan}


\date{\today}

\begin{abstract}
$^{139}$La- and $^{55}$Mn-NMR spectra demonstrate that the ground state of the A-site ordered perovskite manganite LaBaMn$_2$O$_6$ is a spatial mixture of the ferromagnetic (FM) and antiferromagnetic (AFI(CE)) regions, which are assigned to the metallic and the insulating charge ordered state, respectively.
This exotic coexisting state appears below 200 K via a first-order-like formation of the AFI(CE) state inside the FM one.
Mn spin-spin relaxation rate indicates that the FM region coexisting with the AFI(CE) one in LaBaMn$_2$O$_6$ is identical to the bulk FM phase of the disordered form La$_{0.5}$Ba$_{0.5}$MnO$_3$ in spite of the absence of A-site disorder.
This suggests mesoscopic rather than nanoscopic nature of FM region in LaBaMn$_2$O$_6$\@.
\end{abstract}

\maketitle

\sloppy

The perovskite manganites $R_{1-x}A_x$MnO$_3$ ($R$ = rare earth, $A$ = Ca, Ba, Sr) have been attracting much attention for several decades, because of their rich and intriguing electromagnetic properties, such as colossal magnetoresistance (CMR), charge/orbital ordering, and metal-insulator transition.
The CMR effect is believed to be a consequence of competition between double-exchange ferromagnetic metal and superexchange antiferromagnetic insulating phases.
However, the estimated magnetoresistive response by the double-exchange model disagrees with the experimental data by an order of magnitude or more, suggesting the importance of another additional mechanism \cite{millis95}.
This discrepancy may be resolved by considering a phenomenon of phase separation, where the conduction path dominating the resistance depends on the pattern of the coexisting metallic and insulating regions \cite{moreo99,uehara99}.
An external magnetic field may change this pattern, and hence cause a large change in the resistivity.
The phase separation is, thus, an important aspect of manganites and may be an intrinsic feature in many systems with strongly correlated electrons.

Recently, half-doped manganite perovskites $R$BaMn$_2$O$_6$ with the A-site order have been attracting growing interest, because the ordering of $R$ and Ba elements at the A-site of perovskite structure dramatically modifies their phase diagram \cite{millange98,nakajima02,akahoshi03}.
Nakajima {\it et al.}\ have reported that the charge ordering transition temperature is as high as 500 K with a new stacking variation with a fourfold periodicity along the $c$-axis for YBaMn$_2$O$_6$\@ \cite{nakajima02, kageyama03}.
In addition, it is important to make clear whether the disorder at the A site plays a vital role in the occurrence of CMR or not \cite{akahoshi03}.
Hence, new experimental and theoretical works have been devoted to the A-site ordered perovskite manganites to elucidate the effects of A-site order/disorder on the electromagnetic properties \cite{arima02,uchida02,motome03}.

In this paper, we focus our attention on the A-site ordered LaBaMn$_2$O$_6$ and the disordered La$_{0.5}$Ba$_{0.5}$MnO$_3$ with the same composition.
The La and Ba elements are randomly distributed at the A site of perovskite structure in La$_{0.5}$Ba$_{0.5}$MnO$_3$\@.
On the other hand, the structural feature of LaBaMn$_2$O$_6$ is the alternating stack of LaO and BaO layers along the $c$ axis with intervening MnO$_2$ layer.
The ground state of La$_{0.5}$Ba$_{0.5}$MnO$_3$ is a bulk double-exchange ferromagnetic metal (FM) phase with Curie temperature, $T_{\rm C}$ = 280 K \cite{nakajima04}.
LaBaMn$_2$O$_6$ shows a transition from a paramagnetic to a FM phase at $T_{\rm C}$ = 330 K\@.
Remarkably, it is found by the neutron diffraction measurements \cite{nakajima03} that a part of the FM phase transforms to antiferromagnetic charge-exchange-type charge/orbital ordered (AFI(CE)) phase below 200 K and the AFI(CE) phase coexists with the FM phase as the ground state.
It is notable that the observation of phase separation in LaBaMn$_2$O$_6$ suggests that a phase separation is not due to the disorder at A-site but an intrinsic phenomenon in the perovskite manganites.
It is, therefore, crucial to elucidate the nature of electromagnetic properties including phase separation unaffected by the A-site disorder from the microscopic viewpoint with NMR\@.

We have performed $^{139}$La- and $^{55}$Mn-NMR measurements on LaBaMn$_2$O$_6$ and La$_{0.5}$Ba$_{0.5}$MnO$_3$\@.
Both polycrystalline samples were prepared by a solid-state reaction of La$_2$O$_3$, BaCO$_3$, and MnO$_2$\@.
After repeating the sintering process in an Ar flow, the obtained ceramics was annealed in flowing O$_2$, which leads to LaBaMn$_2$O$_6$\@.
The use of pure Ar gas is to avoid the formation of the A-site disorder.
La$_{0.5}$Ba$_{0.5}$MnO$_3$ was obtained by a solid state reaction in 1\% O$_2$/Ar gas, followed by an annealing in O$_2$ gas.
The details are described elsewhere \cite{nakajima04}.
Both samples were characterized crystallographically with an X-ray diffractometer, and magnetically with a SQUID magnetometer.
The obtained data were found to be in accordance with those in the literature \cite{nakajima03}.
NMR was performed with a phase-coherent spectrometer by using a spin-echo technique.

\begin{figure}[t]
\includegraphics[scale=0.27]{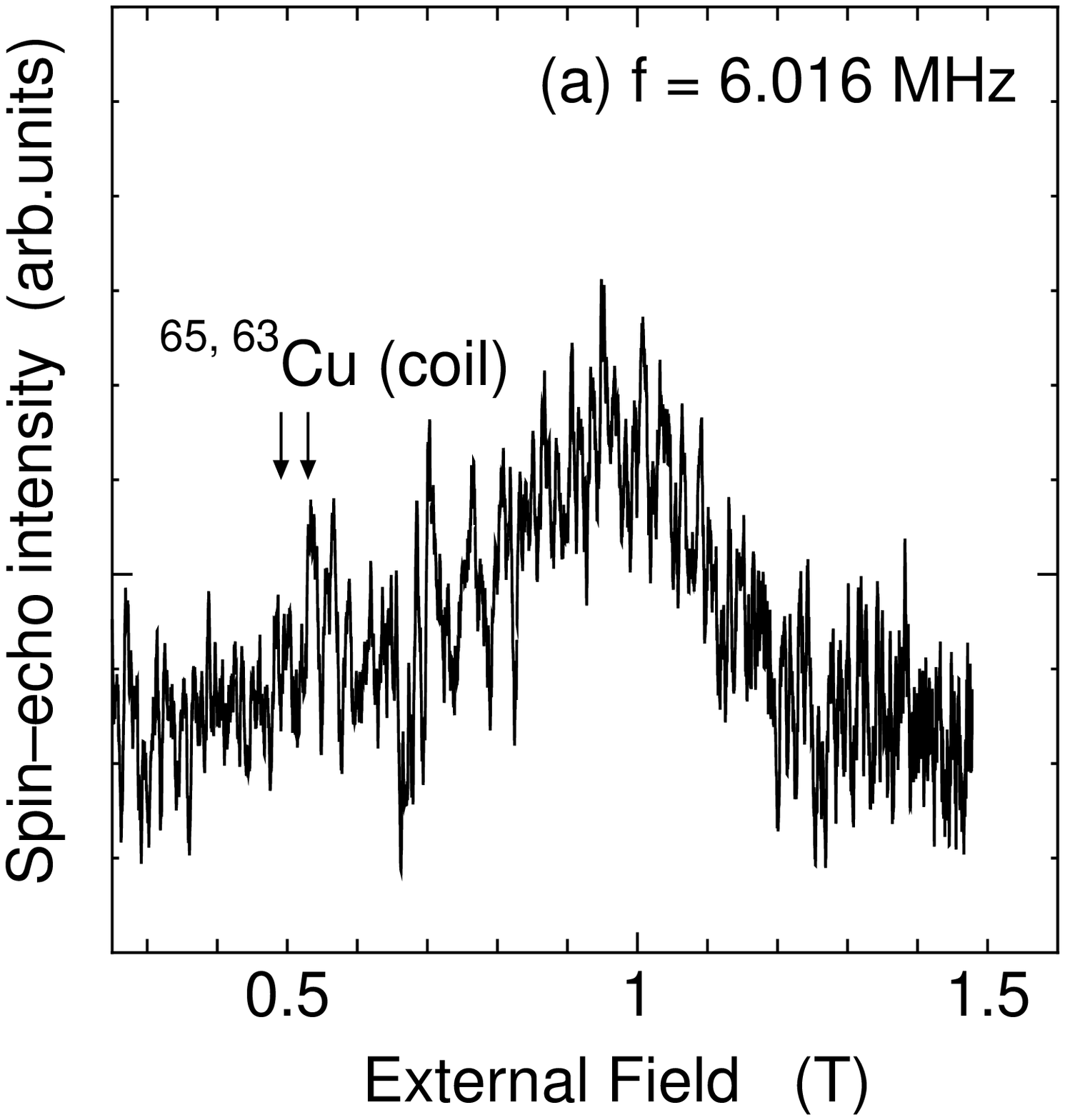}%
\includegraphics[scale=0.27]{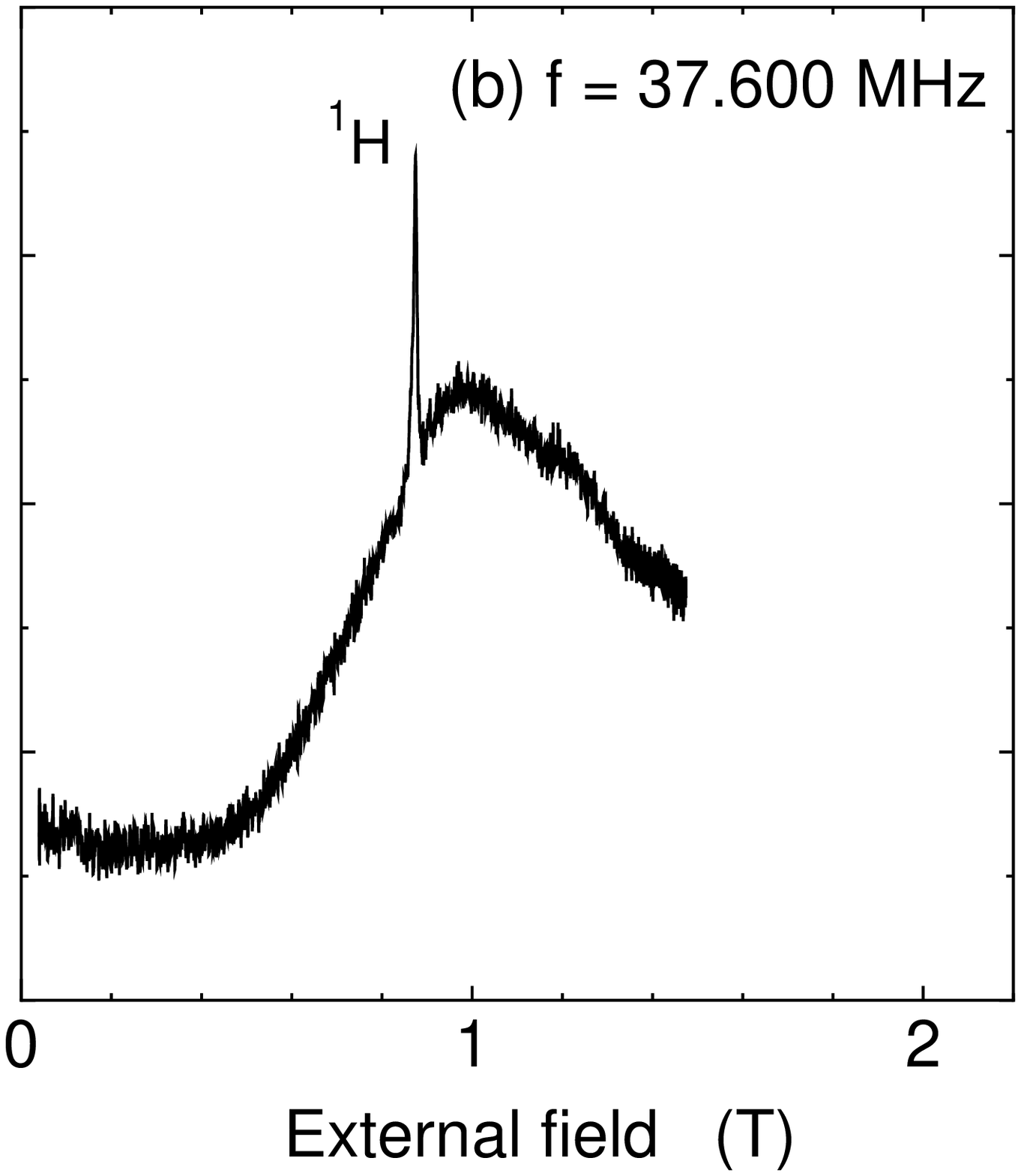}%
\caption{Field-swept $^{139}$La-NMR spectra of LaBaMn$_2$O$_6$ at 1.6 K with (a) $f_1$ = 6.016 MHz and (b) $f_2$ = 37.600 MHz\@.}
\end{figure}

We can easily distinguish between magnetic resonances of nuclei in the FM and AFI(CE) regions.
First, the resonance in the FM region is characterized by a very strong NMR signal and a small applied rf field required for the maximum echo signal due to the coupling between nuclear and electronic spins.
The second distinction between these resonances arises from the different local field at La and Mn sites due to Mn spin and charge order.
The La-NMR signal from the AFI(CE) region is not observable in zero field, because the transferred field from the nearest Mn neighbors cancels at La site in this spin structure \cite{kapusta00, note:dipole}.
On Mn site, previous works in manganese perovskites \cite{papavassilou00} have shown that zero-field antiferromagnetic resonances at the localized Mn$^{4+}$ and Mn$^{3+}$ are around 300-320 MHz and 400-420 MHz, respectively, whereas the signal from FM region is located at an intermediate frequency due to the fast electron transfer between Mn$^{3+}$ and Mn$^{4+}$ ions.

\begin{figure}[t]
\includegraphics[scale=0.43]{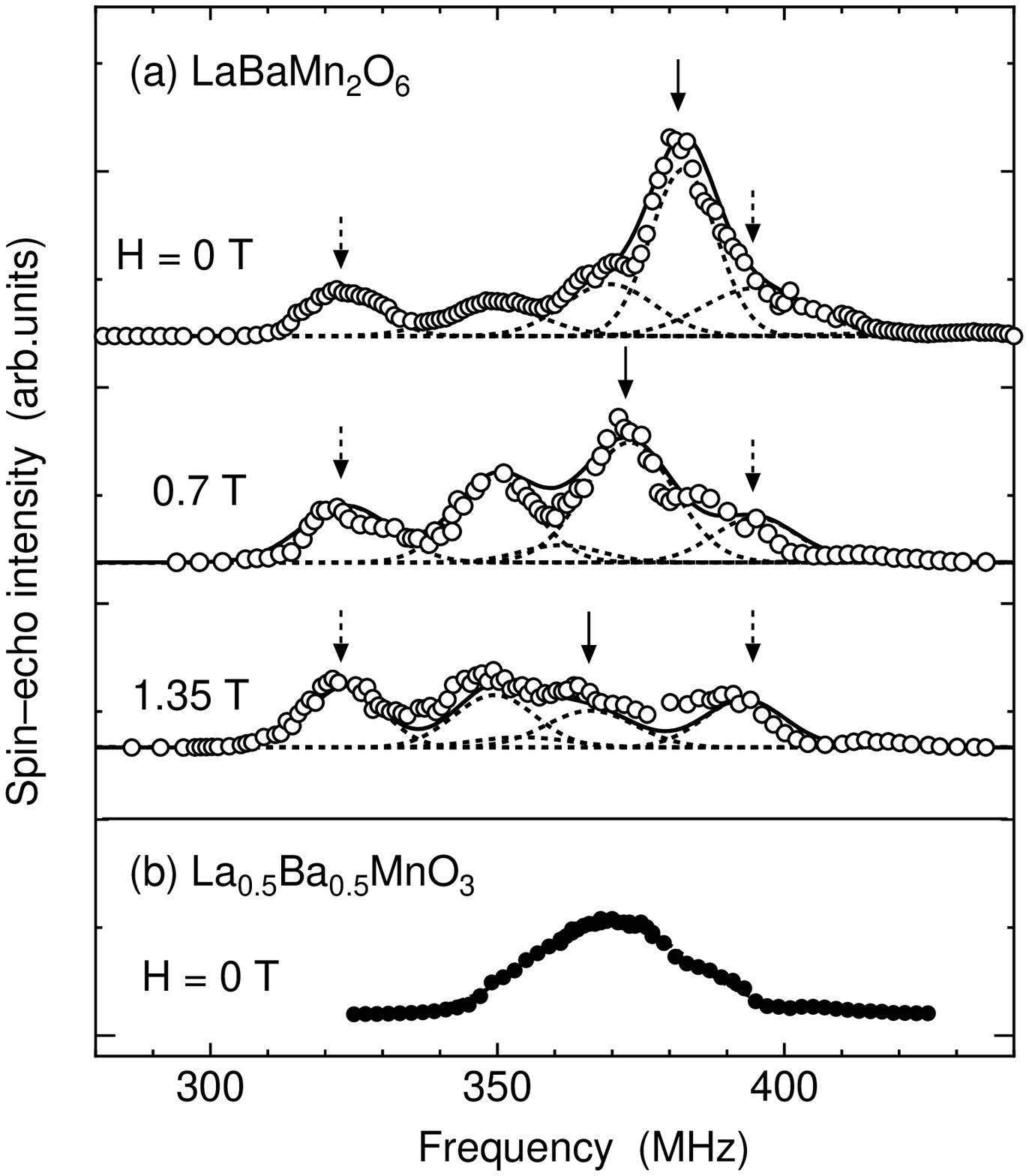}%
\caption{$^{55}$Mn-NMR spectra of (a) LaBaMn$_2$O$_6$ under $H$ = 0, 0.7, and 1.35 T and of (b) \disord\  in zero field at 4.2 K\@.}
\end{figure}

Figure 1 shows the field-swept $^{139}$La-NMR spectra of LaBaMn$_2$O$_6$ at the fixed frequencies of (a) $f_1 =$ 6.016 MHz and (b) $f_2 =$ 37.600 MHz at 1.6 K\@.
The signal around 1.0 T in Fig.~1 (a) is attributed to the resonance from the antiferromagnetic region, because this resonance occurs close to $f_1/^{139}\gamma$ = 1.0 T, namely the hyperfine field cancels at La site in this environment.
On the other hand, the signal around 1.0 T in Fig.~1 (b) is associated with the resonance from the ferromagnetic region with the hyperfine field of 5.2 T via positive transferred couplings.
These assignments are assured by the enhancement observed in the spin-echo intensity only for the signal in Fig.~1 (b).
Thus, these $^{139}$La-NMR spectra give evidence for the coexistence of ferromagnetic and antiferromagnetic regions at the ground state.

$^{55}$Mn-NMR spectra show this coexistence as well.
In Fig.~2 are shown the $^{55}$Mn-NMR spectra of LaBaMn$_2$O$_6$ at $H$ = 0, 0.7 and 1.35 T, and (b) that of La$_{0.5}$Ba$_{0.5}$MnO$_3$ in zero field at 4.2 K\@.
The peak at 370 MHz for LaBaMn$_2$O$_6$ in zero field may be attributed to some region with the disordered structure, because the spectrum for La$_{0.5}$Ba$_{0.5}$MnO$_3$ has a maximum at the same frequency with a ferromagnetic enhancement.
The degree $S$ of A-site order in LaBaMn$_2$O$_6$ is roughly estimated to be about $S$ = 92 \% from the spin-echo intensity, which is comparable to $S = 96\pm2$ \% for PrBaMn$_2$O$_6$ estimated by the X-ray and neutron diffractions \cite{nakajima04, note:degree}.
Here, we use the fact that the FM region occupies about 50 \% of the whole volume of LaBaMn$_2$O$_6$ at 4.2 K as estimated later.
It is also assumed that the enhancement factors are comparable between the resonance for La$_{0.5}$Ba$_{0.5}$MnO$_3$ and that for FM region in LaBaMn$_2$O$_6$\@.

The majority of the signal is associated with the intrinsic LaBaMn$_2$O$_6$ and the origin of their peaks may be distinguished by their shifts in an external field.
The largest peak at 380 MHz (indicated by a solid arrow) is associated with the FM region, ensured by the observation that an external field $H$ shifts this resonance frequency by $-^{55}\gamma H$ and depresses the ferromagnetic enhancement by removing the domain wall that has the largest contribution to the enhancement.
Comparison with the NMR spectra of the related manganites \cite{papavassilou00} indicates that the peaks at 323 MHz and 395 MHz (indicated by dotted arrows) are ascribed to the nuclei of localized Mn$^{4+}$ and Mn$^{3+}$ ions in the AFI(CE) state, respectively.
This is assured by the fact that these peaks are unshifted by an external field.
Thus, both $^{139}$La- and $^{55}$Mn-NMR spectra provide evidence for the coexistence of antiferromagnetic and ferromagnetic regions at the ground state in spite of the ordering of La and Ba, and these regions are assigned to the insulating charge ordered and the metallic states, respectively.
However, the origin of the peak at 349 MHz is unclear at present.
Note that similar additional lines are also observed in the AFI(CE) phase of the A-site ordered YBaMn$_2$O$_6$ \cite{ohno05}.
It may be, therefore, related to the layer-type crystallographical and/or spin structures with a fourfold periodicity along the $c$ axis \cite{kageyama03}.

\begin{figure}[t]
\includegraphics[scale=0.43]{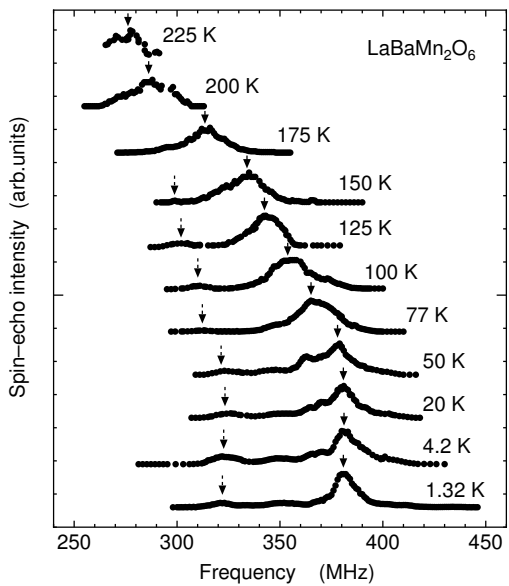}%
\caption{$^{55}$Mn-NMR spectra of LaBaMn$_2$O$_6$ in zero field at several temperatures.}
\end{figure}

Next, we demonstrate the development of the coexisting state as temperature varies.
Figure 3 shows the temperature dependence of the $^{55}$Mn-NMR spectrum for LaBaMn$_2$O$_6$ in zero field.
All spectra are measured at a given temperature after warming from the lowest temperature, otherwise a hysteretic behavior may be observed in the NMR spectra.
Although the line for FM region is observed in the whole measured temperature range, the lines for AFI(CE) region are undetected above 175 K, which is consistent with the appearance of AFI(CE) phase below 200 K as shown later.
The peak frequencies for the FM and AFI(CE) regions, which is proportional to the time-averaged Mn local moments $\mu_{\rm Mn}$, decrease monotonically with increasing temperature.

\begin{figure}[t]
\includegraphics[scale=0.43]{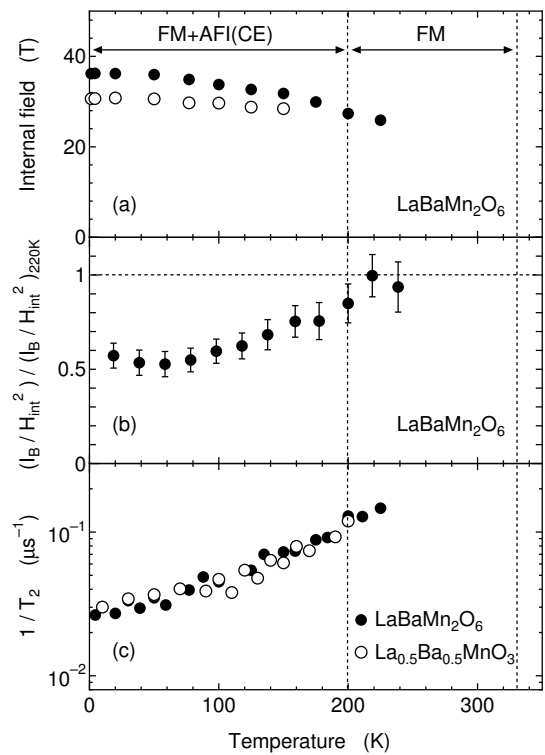}%
\caption{Temperature dependences of (a) $H_{\rm int}^{\rm F}$ (closed circles) and $H_{\rm int}^{\rm AF}$ (open squares), (b) $I_{\rm B}/(H_{\rm int}^{\rm F})^2$ normalized by the value at 220 K, and (c) $1/T_2$'s measured at the peak from the FM region of LaBaMn$_2$O$_6$ (closed circles) and that of La$_{0.5}$Ba$_{0.5}$MnO$_3$ (open circles).
}
\end{figure}

In Fig.~4 (a) is shown the temperature dependence of the internal field estimated from the peak frequency, where closed circles and open squares correspond to the lines from nuclei in the FM region and those of Mn$^{4+}$ in the AFI(CE) region, respectively.
The internal field $H_{\rm int}^{\rm F}$ at Mn site in the FM region decreases monotonically from 36.2 T at 4.2 K with increasing temperature and seems to approach zero at $T_{\rm C} \sim$ 330 K\@.
This temperature dependence seems to be a characteristic of a second-order phase transition, as contrasted with the case in La$_{0.5}$Ca$_{0.5}$MnO$_3$ and Pr$_{0.7}$Ba$_{0.3}$MnO$_3$ exhibiting a first-order-like transition from FM to paramagnetic metal \cite{allodi98,savosta97}.
Note that any anomaly in $H_{\rm int}^{\rm F}$ is not observed around 200 K below which the AFI(CE) region appears inside the FM phase.
This is presumably because the typical size of FM region is mesoscale rather than nanoscale as suggested later and, hence $H_{\rm int}^{\rm F}$ dominated by the on-site core polarization is hardly affected by the spin rearrangement in the region where the FM phase transforms to the AFI(CE) one.
On the other hand, the internal field $H_{\rm int}^{\rm AF}$ in the AFI(CE) region, being still high at 150 K, indicates that the antiferromagnetic spin structure arises from the ferromagnetic one through a first-order-like transition.
In connection with this fact, the peak frequency 349 MHz at 4.2 K becomes close to that for the FM region with increasing temperature, which may be responsible for the apparent broadening of the NMR line from the FM region above 77 K\@.

To gain further insight into the phase separation, we compare our results with those obtained by the neutron diffraction experiments \cite{nakajima03}.
While $H_{\rm int}^{\rm F}$ is proportional to $\mu_{\rm Mn}$ in the FM region, the magnetic Bragg neutron-reflection intensity $I_{\rm B}$ has a relation $I_{\rm B} \propto (V_{\rm FM}/V_0) \mu_{\rm Mn}^2$, where $V_{\rm FM}$ and $V_0$ are the volumes of the FM region and the whole sample, respectively.
The value of $I_{\rm B}/(H_{\rm int}^{\rm F})^2$ is, therefore, proportional to the FM volume fraction, $V_{\rm FM}/V_0$\@.
Figure 4 (b) shows the temperature dependence of $I_{\rm B}/(H_{\rm int}^{\rm F})^2$ normalized by the value at 220 K \cite{note:enhance}.
The FM volume fraction decreases below 200 K and remains about 50\% below 100 K, which suggests a formation of the AFI(CE) phase inside of the FM phase.
The estimated FM volume fraction is somewhat smaller than 70 \% reported from the magnetization \cite{nakajima03}.
The bulk ferromagnet La$_{0.5}$Ba$_{0.5}$MnO$_3$ contained in LaBaMn$_2$O$_6$ may be responsible for this discrepancy.

Finally, the temperature dependence of Mn spin-spin relaxation rate $1/T_2$ is shown in Fig.~4 (c), where closed and open circles are the data measured at the peaks from the nuclei in the FM region of LaBaMn$_2$O$_6$ and those of La$_{0.5}$Ba$_{0.5}$MnO$_3$, respectively.
$1/T_2$'s are determined by a fit with a relaxation $I(\tau)=I(0) \exp(-2\tau/T_2)$, where $I(\tau)$ is a spin-echo intensity at a pulse interval $\tau$ in the spin-echo measurements.
They show almost the same temperature dependence, that is, an exponential rise $1/T_2 \propto \exp(aT)$ in a wide temperature range above 60 K\@.
Such a temperature variation in $1/T_2$, which is rather different from the case in a conventional ferromagnetic metal, has been reported for a number of double-exchange FM manganites \cite{savosta99} including La$_{0.5}$Ba$_{0.5}$MnO$_3$\@.
This result indicates that the nature of FM region coexisting with the AFI(CE) one in LaBaMn$_2$O$_6$ is microscopically identical to the bulk FM phase of La$_{0.5}$Ba$_{0.5}$MnO$_3$, irrespective of the order/disorder at the A-site of perovskite.
In addition, this suggests mesoscopic rather than nanoscopic nature of FM region in LaBaMn$_2$O$_6$\@.
The small deviation from the exponential rise for both samples below 60 K may be ascribed to an additional relaxation process, such as the Suhl-Nakamura interaction in which the nuclear spins interact via electronic spin waves \cite{savosta99,suhl58,nakamura58}.
No indication of critical behavior around 200 K in $1/T_2$ for LaBaMn$_2$O$_6$ may be related to the first-order-like formation of the AFI(CE) phase inside the FM phase.

In conclusion, it is demonstrated by the NMR spectra that the FM and AFI(CE) regions coexist at the ground state of LaBaMn$_2$O$_6$ in spite of the A-site order, where the AFI(CE) region occupies about half volume of the whole sample.
This exotic coexisting state appears below 200 K via a first-order-like formation of the AFI(CE) state inside the FM one.
The temperature dependence of $1/T_2$ indicates that the FM region coexisting with the AFI(CE) one in LaBaMn$_2$O$_6$ is microscopically identical to the bulk FM phase of La$_{0.5}$Ba$_{0.5}$MnO$_3$, irrespective of the order/disorder at the A-site of perovskite.
This suggests mesoscopic rather than nanoscopic nature of FM region in LaBaMn$_2$O$_6$\@.


This work was in part supported by Grant-in-Aid for Scientific Research (Nos.\ 16540324 and 17740231) from MEXT of Japan and by Yazaki Memorial Foundation for Science and Technology\@.

\clearpage
%
%
%
%
%
%

\begin{references}

\bibitem{millis95}
A.J.~Millis, P.B.~Littlewood, and B.I.~Shraiman, Phys.\ Rev.\ Lett., {\bf 74}, 5144 (1995).
\bibitem{moreo99}
A.~Moreo {\it et al.}, Science, {\bf 283}, 2034 (1999).
\bibitem{uehara99}
M.~Uehara, et al., Nature, {\bf 399}, 560 (1999).
\bibitem{millange98}
F.~Millange {\it et al.}, Chem.\ Mager.\ {\bf 10}, 1974 (1998).
\bibitem{nakajima02}
T.~Nakajima {\it et al.}, J.\ Phys.\ Soc.\ Jpn., {\bf 71}, 2843 (2002).
\bibitem{akahoshi03} 
D.~Akahoshi {\it et al.}, Phys.\ Rev.\ Lett., {\bf 90}, 177203 (2003).
\bibitem{kageyama03}
H.\ Kageyama {\it et al.}, J.\ Phys.\ Soc.\ Jpn., {\bf 72}, 241 (2003).
\bibitem{arima02}
T.\ Arima {\it et al.}, Phys.\ Rev.\ B, {\bf 66}, 140408(R) (2002).
\bibitem{uchida02}
M.\ Uchida {\it et al.}, J.\ Phys.\ Soc.\ Jpn., {\bf 71}, 2605 (2002).
\bibitem{motome03}
Y.\ Motome, N.\ Furukawa, and N.\ Nagaosa, Phys.\ Rev.\ Lett., {\bf 91}, 167204 (2003).
\bibitem{nakajima04}
T.\ Nakajima {\it et al.}, J.\ Phys.\ Soc.\ Jpn., {\bf 73}, 2283 (2004).
\bibitem{nakajima03}
T.\ Nakajima {\it et al.}, J.\ Phys.\ Soc.\ Jpn., {\bf 72}, 3237 (2003).
\bibitem{kapusta00}
Cz.\ Kapusta {\it et al.}, Phys.\ Rev.\ Lett., {\bf 84}, 4216 (2000).
\bibitem{note:dipole}
A dipole contribution is estimated to be as negligibly small as a few kG\@.
\bibitem{papavassilou00}
G.\ Papavassiliou {\it et al.}, Phys.\ Rev.\ Lett., {\bf 84}, 761 (2000).
\bibitem{note:degree}
\protect The degree of A-site order is expressed as $S = (2g-1) \times 100$ \% for [$R_g$Ba$_{1-g}$]$_R$[$R$$_{1-g}$Ba$_g$]$_{\rm B}$Mn$_2$O$_6$, where [~]$_R$ (or [~]$_{\rm B}$) represents $R$-sites (or Ba-sites) in $R$BaMn$_2$O$_6$\@.
\bibitem{ohno05}
T.\ Ohno {\it et al.}, Physica B, {\bf 359-361}, 1291 (2005).
\bibitem{allodi98}
G.~Allodi {\it et al.}, Phys.\ Rev.\ Lett., {\bf 81}, 4736 (1998).
\bibitem{savosta97}
M.M.~Savosta {\it et al.}, Phys.\ Rev.\ Lett., {\bf 79}, 4278 (1997).
\bibitem{note:enhance}
It is difficult to estimate volume fraction of each phase from NMR signal intensity, because the temperature dependence of enhancement factor is not clear.
\bibitem{savosta99}
M.M.\ Savosta, V.A.\ Borodin, and P.\ Nov\'ak, Phys.\ Rev.\ B {\bf 59}, 8778 (1999).
\bibitem{suhl58}
H.~Shul, Phys.\ Rev., {\bf 109}, 606 (1958).
\bibitem{nakamura58}
T.~Nakamura, Prog.\ Theor.\ Phys., {\bf 20}, 542 (1958).

\end{references}
\end{document}